\documentclass[twocolumn,amsmath,amssymb]{revtex4}
\usepackage{graphicx}% Include figure files
\usepackage{dcolumn}% Align table columns on decimal point
\usepackage{bm}% bold math
\begin{document}

%\preprint{APS/123-QED}

\title{High Quality factor photonic crystal nanobeam cavities}

\author{Parag B. Deotare, Murray W. McCutcheon, Ian W. Frank, Mughees Khan and Marko Lon\v{c}ar}
\affiliation{School of Engineering and Applied Sciences, Harvard University, Cambridge, MA 02138
}

\date{\today}% It is always \today, today,

\begin{abstract}

We investigate the design, fabrication and experimental characterization of high Quality
factor photonic crystal nanobeam cavities in silicon.  Using a five-hole tapered 1D photonic crystal
mirror and precise control of the cavity length, we designed cavities with theoretical Quality
factors as high as $1.4 \times 10^7$.  By detecting the cross-polarized resonantly scattered
light from a normally incident laser beam, we measure a Quality factor of nearly
$7.5 \times 10^5$.  The effect of cavity size on mode frequency and Quality factor was
simulated and then verified experimentally.

\end{abstract}

\maketitle

For the past decade, there has been a concerted research effort to develop ultra-high Quality
($Q$) factor electromagnetic cavities with dimensions comparable to the wavelength of
light~\cite{Painter99,Noda03, Noda05, Kuramochi06, Srinivasan_OE_04, Topolancik}.
By shrinking the modal volume to near the fundamental limit of $V = (\lambda/2n)^3$,
these cavities have enabled new applications to emerge in
ultrasmall lasers~\cite{PainterScience, loncar_apl_2004, Baba, Park_04}, strong light-matter
coupling~\cite{Yoshie04,Reithmaier, Hennessy07, Englund07,Srinivasan07,VuckovicPRE},
optical switching~\cite{Tanabe05}, and chemical sensing~\cite{Loncar, erickson}, among others. 
Recently, there has been much interest in cavities realized in
suspended nanobeams patterned with a one-dimensional (1D) lattice of
holes~\cite{Sauvan,Zain_08,McCutcheon_08, Notomi_1D} due to their exceptional cavity figures of
merit ($Q$ and $V$), relative ease of design and fabrication, and potential for novel
optomechanical effects~\cite{Povinelli,Eichenfeld_08}. These apparently simple structures, which
resemble very early microcavity prototypes~\cite{Foresi_97}, actually have $Q/V$ factors which
rival the best 2D planar photonic crystal cavities~\cite{Noda05,Kuramochi06}.  They also have
many inherent advantages, including the possibility of realizing high $Q/V$ cavities in 
moderate index materials such as SiN$_x$~\cite{McCutcheon_08} and
facilitating coupling to ridge waveguides~\cite{Zain_08}.
In addition, the near-field of the cavity is also highly ``accessible'', in the sense that there 
are two dimensions with total-internal-reflection (TIR) interfaces, which should facilitate
bio-sensing applications as well as novel techniques for the dynamic control of cavity
resonances~\cite{Koenderink_PRL}.

\begin{figure}[b]
\begin{center}
\includegraphics[width=7cm]{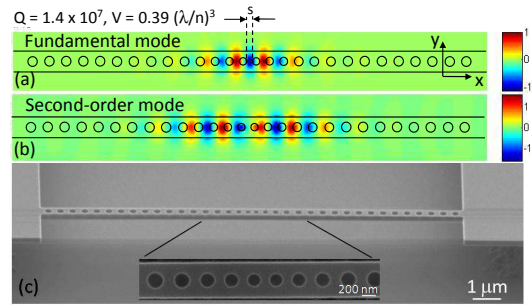}
\end{center}
\vspace{-10pt}
\caption{(a,b) Electric field ($E_y$) profiles of the two cavity modes. $Q$ and $V$
are quoted for the optimal cavity length, $s$. (c) SEM image of a fabricated photonic
crystal nanobeam cavity.  The nanobeam thickness is 220 nm and width is 500 nm.
 The photonic mirror pitch $a$ = 430 nm is linearly tapered over a 5 hole section to
$a$ = 330 nm at the cavity center.  The hole radius is given by r = 0.28a. The 1D photonic bandgap extends from 1200-1700 nm.
% photonic bandgap depends on pitch and this is the only place where we talk about the %pitch, so I moved that info here
\label{fig:modes}}
\end{figure}

In this paper we describe the
design, fabrication, and experimental characterization of silicon photonic crystal nanobeam (PhCnB)
cavities operating near $\sim1500$ nm with measured $Q$ factors of 7.5 $\times 10^5$. To our
knowledge, this represents the highest $Q$ factor ever measured in nanocavities based on
photonic crystal nanobeams, and one of the highest $Q$s ever measured in any photonic crystal cavity.
Electromagnetic field confinement in the structure [Fig.~\ref{fig:modes}(a)] is achieved
by index guiding in two directions ($y$ and $z$), and Bragg scattering from the 1D photonic crystal
mirror in the third ($x$) direction.  The mechanism of light confinement has been
interpreted in terms of impedance matching~\cite{Sauvan, Lalanne_JQE, McCutcheon_08} and the
mode-gap effect~\cite{Notomi_1D}. Conceptually, the cavity can be viewed as a wavelength-scale
Fabry-Perot cavity with photonic crystal mirrors which reflect and thus trap the nanobeam waveguide
mode.  Because the cavity mode penetrates some distance into the mirror, it is crucial that the
fields do not abruptly terminate at the mirror boundary, as this would lead to considerable
scattering loss~\cite{Lalanne_JQE}.
To avoid this impedance mismatch between the waveguide mode and the Bloch mode,
the photonic crystal mirror is tapered by reducing the hole spacing ($a$) and radius to match the
effective indices of the evanescent mirror Bloch mode, $n_{\rm Bl} = \lambda/2a$, and the waveguide
mode, $n_{\rm wg}=2.41$.  The cavities were designed using the 3D finite-difference time-domain (FDTD)
method (Lumerical Solutions, Inc.) and incorporate a five-hole linear taper in a free-standing
silicon nanobeam of thickness 220 nm (constrained by our experimental wafer) and width 500 nm, as
detailed in Fig. 1. With the
tapered mirrors designed to minimize reflection loss from the incident lowest-order waveguide
mode, the cavity length is scanned to optimize the Quality factor of the fundamental cavity
mode, as shown in Fig. ~\ref{fig:spec}(c).  The $Q$ factor is calculated from the definition:
$Q = \omega_0 \frac{\rm Energy\;stored}{\rm Power\;loss}$, and is validated
in the lower $Q$ structures by monitoring the time-domain ring-down of the fields.
The optimal structure supports a fundamental mode at $\lambda=1560$ nm with a 
$Q$ factor of $1.4 \times 10^7$ and
an ultra-small mode volume of V = 0.39 ($\lambda$/n)$^3$.  The mode profile is plotted in
Fig.~\ref{fig:modes}(a).  The cavity also supports higher-order modes
with different symmetry, one of which is shown in Fig.~\ref{fig:modes}(b), and
which has a reasonable $Q$ factor of 120,000 and mode volume  V = 0.71 ($\lambda$/n)$^3$.

%%%%%%%%%%%%%%%%% figure %%%%%%%%%%%%%%%%%%
%%%%%%%%%%%%%%%%%%%%%%%%%%%%%%%%%%%%%%%%%%%%

The devices were fabricated on a silicon-on-insulator (SOI) substrate (SOITEC Inc) with a
device layer of 220nm and an insulator layer of 2 $\mu$m.  A negative electron-beam (e-beam)
lithography resist, Foxx-17 (Dow Corning) diluted in Anisole in a 1:6 ratio, was used for e-beam
lithography. The film was spin coated onto the sample at 5000 rpm and then baked at 90$^\circ$C on
a hot plate for 5 minutes, resulting in a total film thickness of 135 nm.  Patterns were defined
using a standard 100kV e-beam lithography tool (Elionix) and developed in tetra-methyl ammonium
hydroxide (25\% TMAH) followed by a thorough DI water rinse. The devices were etched in a reactive
ion etcher (STS-ICP RIE) using SF$_6$, C$_4$F$_8$ and H$_2$ gases. Removal of the oxide sacrificial
layer was carried out using a HF vapor etching (HFVE) tool (AMMT) operating at 35$^\circ$C, which
resulted in an etch rate of approximately 125 nm/min~\cite{deotare_hfvapor_2009}. 
This technique produced more reliable results than the more conventional approach of a wet etch 
followed by critical point drying. A complete fabricated device is shown in
Fig.~\ref{fig:modes}(c).  A number of photonic crystal cavities were made by systematically varying
the cavity length (s), resulting in resonators with different Quality factors and operating wavelengths.

%%%%%%%%%%%%%%%%% figure %%%%%%%%%%%%%%%%%%
\begin{figure}[b]
\begin{center}
\includegraphics[width=5cm]{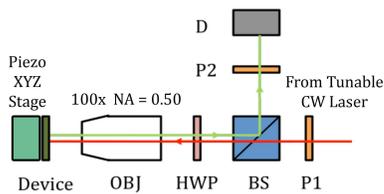}
\end{center}
\vspace{-10pt}
\caption{Schematic of experimental set-up (OBJ: microscope objective, HWP: half-wave plate,
BS: beamsplitter, P1 and P2: polarizers, D: detector).\label{fig:rs}}
\end{figure}
%%%%%%%%%%%%%%%%%%%%%%%%%%%%%%%%%%%%%%%%%%%%

The fabricated devices were tested using a resonant scattering optical setup~\cite{McCutcheon_05,Rivoire}
(Fig.~\ref{fig:rs}).
Prior to entering the objective lens, the polarization of the incident laser beam is rotated by
45$^\circ$ using a half-wave plate (HWP), so that the E-field of the focused laser spot and the major
component of the cavity mode ($E_y$) form a 45$^\circ$ angle.  Light coupled in and subsequently
re-emitted (back-scattered) by the cavity is collected using the same objective lens, and then its
polarization is rotated by -45$^\circ$ after passing through the same HWP. The back-scattered signal
is then split using a beam splitter (BS), analyzed using a polarizer (P2) that is cross-polarized with
respect to the polarization of the incoming laser beam, and finally detected using an InGaAs detector.
This combination of polarizers and wave plates enhances the ratio between the resonantly scattered
signal from the cavity and the coherent background due to non-resonant reflections.  The spectra are
normalized by a non-resonant background taken away from the cavity.  Because of the coherent
relationship between the scattered signal and the background, resonances appear as dips or peaks (depending on
the geometry) and can have an asymmetric shape similar to Fano resonances.
In this scheme, the cavity plays the role of a wavelength selective polarization
rotator.  Our experimental
approach therefore allows for resonant spectroscopy of the cavity, and does not require integration
of additional waveguides to couple light in and out of the cavity.  Therefore, our method measures
the intrinsic $Q$ factor of the cavity without ``loading'' effects due to the presence of coupling
waveguides.

%%%%%%%%%%%%%%%%% figure %%%%%%%%%%%%%%%%%%
\begin{figure}[t]
\begin{center}
\includegraphics[width=7cm]{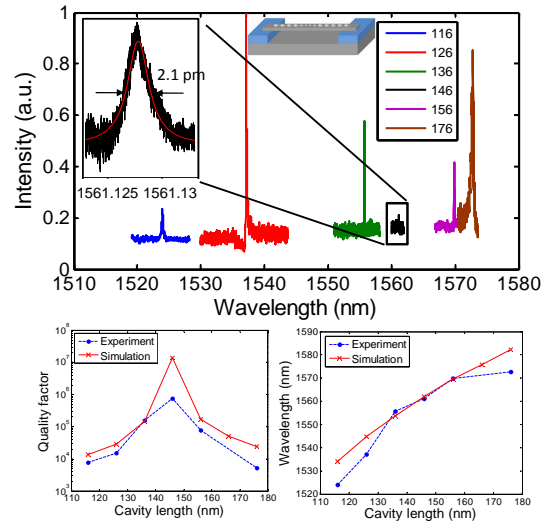}
\end{center}
\vspace{-10pt}
\caption{(a) Resonant scattering spectra for a range of cavities (s = 116-176 nm)
normalized by a background spectrum taken on the beam away from
the cavity. (b) Measured and simulated
Quality ($Q$) factors as a function of cavity length (s).(c) Mode resonance wavelength as a function of s.
\label{fig:spec}}
\end{figure}
%%%%%%%%%%%%%%%%%%%%%%%%%%%%%%%%%%%%%%%%%%%%

The experimental results are shown in Fig.~\ref{fig:spec}.  We expect that as the cavity becomes
longer, the resonance should redshift due to the increase in the effective index of the cavity.
This behavior is well modelled by our simulations, and we see a similar overall trend in our
experimental results. Detailed investigation of fabricated structures using scanning electron
microscopy (SEM) revealed that deviations from theoretical results can be attributed to
fabrication-related disorders, and in particular to proximity effects during e-beam lithography.
The cavity $Q$ factor also follows the predicted trend.  As expected, for large and
small $s$, the $Q$ factor is modest ($Q<10,000$), and it reaches a record high value of $Q=750,000$
when $s=146$ nm. We also found that the cavity
resonance was very sensitive to the excitation location, and disappeared with sub-micron displacements
of the cavity. This is consistent with the expected small mode volume of our cavities.

We fabricated a range of structures with scaled dimensions to account for imperfections introduced
during fabrication 
and to effectively bracket the design parameters.  We found that the structures scaled by -3\% most
closely matched our simulations, and the spectra from these cavities are the ones
presented in Fig.~\ref{fig:rs}.  It is interesting to note that cavities with a higher $Q$
factor were more difficult to characterize using our resonant scattering setup, and had a reduced
contrast between the resonant feature and the coherent non-resonant background. This may be an intrinsic 
property and fundamental limitation of the resonant scattering approach, and it will be addressed in our 
future publications.

%%%%%%%%%%%%%%%%% figure %%%%%%%%%%%%%%%%%%
\begin{figure}[t]
\begin{center}
\includegraphics[width=7cm]{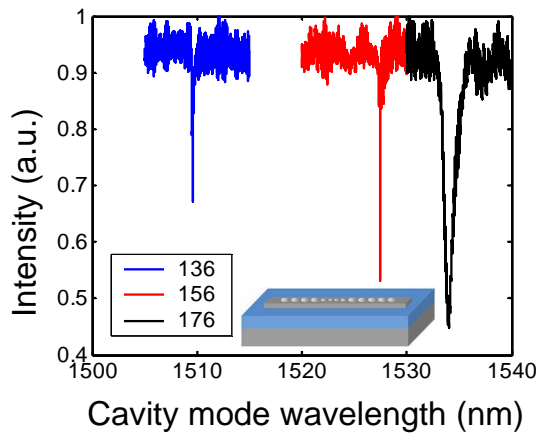}
\end{center}
\vspace{-10pt}
\caption{Resonant scattering spectra for on-substrate SOI cavities (not undercut).
The $Q$ factors are 18,000; 27,500; and 17,000
for the $s$ = 136, 156, and 176 nm cavities, respectively.
\label{fig:oxide}}
\end{figure}
%%%%%%%%%%%%%%%%%%%%%%%%%%%%%%%%%%%%%%%%%%%%

We also tested our nano-beam resonators before the final release HFVE step. Non-suspended,
on-substrate cavities are more robust and are suitable for such applications as sensing and operation 
in fluids~\cite{erickson}.  Fig.~\ref{fig:oxide} shows the experimental results for
several resonator designs with varying cavity spacing, $s$, taken before the cavities were released
from the substrate. The highest $Q$ factor that we were able to obtain was
27,500 for a cavity with $s=156$ nm.  Higher $Q$ factor results have been
previously reported in different on-substrate PhCnB cavities~\cite{Zain_08}. We note, however, that
our cavities were optimized for free-standing operation, and therefore the modest $Q$ factors are
not surprising.

In conclusion, we have successfully designed and fabricated ultra high-$Q$ photonic crystal nanobeam cavities 
using our novel five-hole taper design with a measured $Q$ factor of $7.5 \times 10^5$. The
devices on silicon dioxide have also been studied and show a moderate $Q$ factor, on the order of $10^4$.
We have also reported, for the first time, the successful use of the resonant scattering method for
measuring photonic crystal nanobeam cavities.

This work is supported in part by NSF ECCS-0701417 and NSF CAREER grants.  MWM would like to thank
NSERC (Canada) for its support. IWF acknowledges support by the NSF graduate student fellowship. 
Device fabrication was performed at the Center for Nanoscale
Systems at Harvard.
\vspace{-20pt}
%\bibliography{1dpaper}% Produces the bibliography via BibTeX.

%%%%%%%%%%%%%%%%% figure %%%%%%%%%%%%%%%%%%
%\begin{figure*}[htb]
%\begin{center}
%\vspace{10pt}
%\includegraphics[width=10cm]{nonlinearb.eps}
%\end{center}
%\caption{\label{fig:nonlinear}}
%\end{figure*}
%%%%%%%%%%%%%%%%%%%%%%%%%%%%%%%%%%%%%%%%%%%%

%\pagebreak

\end{document}